\documentclass[fleqn,usenatbib]{mnras}

\usepackage{newtxtext,newtxmath}
\usepackage{graphicx}
\usepackage{subcaption}

\usepackage[T1]{fontenc}
\usepackage{bm}
\usepackage{amsmath}
\usepackage{balance}

\DeclareRobustCommand{\VAN}[3]{#2}
\let\VANthebibliography\thebibliography
\def\thebibliography{\DeclareRobustCommand{\VAN}[3]{##3}\VANthebibliography}

\usepackage{graphicx}	
\usepackage{amsmath}	

\usepackage[usenames,dvipsnames]{xcolor}

\usepackage{orcidlink}
\hypersetup{
    colorlinks = true,
    citecolor = {MidnightBlue},
    linkcolor = {BrickRed},
    urlcolor = {BrickRed}
}

\title[Doppler magnification dipole with LSST]{Prospects for measuring the Doppler magnification dipole with\\
LSST and DESI}

\author[Isabelle Ye et al.]{
Isabelle Ye\,\orcidlink{0009-0007-1958-3364},$^{1}$\thanks{E-mail: isabelle.ye@postgrad.manchester.ac.uk}\
Philip Bull\,\orcidlink{0000-0001-5668-3101},$^{1,2}$\
Caroline Guandalin\,\orcidlink{0000-0003-1490-9314},$^{3}$\
Chris Clarkson\,\orcidlink{0000-0001-7363-0722},$^{2,4}$\
Ainulnabilah Nasirudin\,\orcidlink{0000-0003-2213-4547}$^{1}$
\\
$^{1}$Jodrell Bank Centre for Astrophysics, University of Manchester, Manchester M13 9PL, UK\\
$^{2}$Department of Physics and Astronomy, University of Western Cape, Cape Town 7535, South Africa\\
$^{3}$Institute for Astronomy, University of Edinburgh,
Royal Observatory, Blackford Hill, Edinburgh EH9 3HJ, UK\\
$^{4}$School of Physical and Chemical Sciences, Queen Mary University of London, London E1 4NS, UK\\
}

\date{Accepted XXX. Received YYY; in original form ZZZ}

\pubyear{2025}

\begin{document}
\label{firstpage}
\pagerange{\pageref{firstpage}--\pageref{lastpage}}
\maketitle

\begin{abstract}
We forecast the detectability of the Doppler magnification dipole with a joint analysis of galaxy spectroscopic redshifts and size measurements. The Doppler magnification arises from an apparent size variation caused by galaxies' peculiar velocities when mapping them from redshift space to real space. This phenomenon is the dominant contribution to the convergence at low redshifts ($\lesssim$ 0.5). A practical observational strategy is to cross-correlate a galaxy number count tracer, e.g. from the Dark Energy Spectroscopic Instrument (DESI) Bright Galaxy Survey, with the convergence field reconstructed from galaxy size measurements obtained by the Vera C. Rubin Observatory's Legacy Survey of Space and Time (LSST). To assess the achievable precision of galaxy size measurements, we simulate LSST Y1-quality galaxy images with \textsc{Galsim} and measure them with the \textsc{Galight} profile fitting package. Our investigations, based on galaxy populations from LSST's synthetic galaxy catalogue \textsc{cosmoDC2}, show that the variance due to intrinsic galaxy size variation dominates over size measurement errors as expected, but may be lower than previous studies have suggested. Under our analysis assumptions, the Doppler magnification dipole would be detectable with a signal-to-noise ratio $\geq 10$ in multiple redshift bins between $0.1 \leq z \leq 0.5$ with DESI spectroscopic redshifts and LSST imaging.
\end{abstract}




\section{Introduction}

Peculiar velocities are the motions of galaxies driven by gravitational forces from nearby structures, separate from the apparent velocity associated with the homogeneous and isotropic cosmic expansion described by Hubble's law. These deviations from the Hubble flow trace the growth of large-scale structure and the underlying matter distribution \citep[e.g.][]{1989ApJ...344....1G, 2010MNRAS.407.2328F, 2014MNRAS.445.4267K, Turner_2024}, and are therefore a valuable cosmological probe that allows us to test General Relativity and its possible alternatives \citep[e.g.][]{2014PhRvL.112v1102H, 2016A&A...595A..40I, 2023MNRAS.518.5929L}. Since a galaxy’s peculiar motion along the line of sight contributes an extra Doppler shift to its observed redshift, measuring this velocity requires an independent estimate of the galaxy’s distance to infer the expected Hubble flow redshift at that location. In practice, this is commonly achieved by using redshift-independent distance indicators such as Type Ia supernovae \citep{Phillips_1993}, scaling relations such as the Tully-Fisher relation \citep{Tully_1977} for spiral galaxies or the Fundamental Plane \citep{Djorgovski_1987,Dressler_1987} for ellipticals, and even gravitational wave events \citep{Holz_2005}. Statistically, in galaxy redshift surveys, it can also be measured by the redshift-space distortion (RSD) through characteristic anisotropies in the clustering of galaxies \citep{Kaiser_1987}.

A less direct but powerful consequence of galaxy peculiar motions is the Doppler magnification effect \citep{Bonvin_2008, Bacon_2014}. It arises from peculiar velocities shifting galaxy locations in redshift space, causing shifts in their apparent flux and angular size as a function of observed redshift. This redshift-size mismatch induces an apparent lensing convergence and magnification signal. At fixed observed redshift, galaxies that are actually moving toward the observer appear fainter and smaller than naively expected, while those receding appear brighter and larger than they would have done if their observed redshift were equal to their true cosmological redshift. Around an overdensity, this generates a dipolar magnification pattern, as near-side infalling galaxies appear magnified and far-side receding ones are demagnified \citep{Bacon_2014}.

Over the past decade, a number of studies have built the theoretical foundation for Doppler magnification as a cosmological probe. The effect was first highlighted by \cite{Bonvin_2008}, who noted that the galaxy's velocity would induce an apparent magnification of its image. Subsequent works showed that at low redshifts, this Doppler-induced magnification can dominate over the gravitational lensing signal \citep{Bacon_2014,Bonvin_2017,Coates_2021,bonvin2023case}. \cite{Bacon_2014} introduced the term ``Doppler lensing'' and used simulations to demonstrate that a high signal-to-noise ratio (SNR) detection could be achieved with upcoming surveys. Based on these insights, \cite{Bonvin_2017} developed a practical estimator to isolate the Doppler magnification dipole signal, by extracting the dipole term in the cross-correlation between galaxy number counts and convergence. This estimator largely cancels symmetric lensing contributions up to $z\sim0.5$, thereby providing a cleaner measurement. The predicted dipole signal is strongest at low redshifts where peculiar velocities can be a significant contribution to the observed redshift. This is a direct way to trace the galaxy peculiar velocity field, and is an independent probe of the growth rate that complements more traditional RSD measurements (e.g., \citealt{Zarrouk_2018,Icaza_2020}).

The number count of galaxies is a well-studied observable \citep{Bonvin_2011,Challinor_2011}, but using galaxy sizes to infer convergence is less explored. Galaxy size estimates are a by-product of weak lensing surveys, which tend to focus on the shear signal by measuring galaxy ellipticities. \cite{Casaponsa_2013} showed that unbiased estimates of the convergence field (and thus cosmic size magnification) can be extracted from galaxy samples provided that the galaxies have angular sizes larger than the point spread function (PSF) and have high SNR ($>10$). Similarly, \cite{Alsing_2015} offered a Bayesian methodology for estimating the convergence field based on galaxy angular size and flux measurements. These studies established the observational viability of using the galaxy distribution and sizes as a probe of lensing convergence (and thus the Doppler magnification signal).

\cite{Andrianomena_2019} presented encouraging forecasts for surveys carried out with the Dark Energy Spectroscopic Instrument \citep[DESI;][]{DESI_2022} and the futuristic Square Kilometre Array (SKA) Phase 2 with the 21cm line, finding that they could detect the Doppler magnification dipole with high SNR. Such measurements could be used to test cosmological models and modifications to General Relativity on large scales, and would provide a useful alternative to other observables as this peculiar velocity lensing probe is sensitive to different systematics than standard methods. More recently, \cite{Coates_2021} employed a fully relativistic $N$-body simulation to verify that the Doppler magnification term in the lensing convergence can be recovered as predicted by relativistic perturbation theory. In addition, \cite{bonvin2023case} split a galaxy sample by luminosity (into ``bright'' and ``faint'' subsamples) and forecast a detectable $\sim20\sigma$ dipole in the two-point correlation function using DESI's Bright Galaxy Survey \citep[BGS;][]{Hahn_2023}.

Galaxy size magnification due to the Doppler effect is a subtle signal, and has not yet been detected. However, with upcoming surveys like the Legacy Survey of Space and Time \citep[LSST;][]{Ivezi_2019} and DESI, it is important to further assess its detectability, while also taking into account realistic observational complications such as the PSF and detector effects. In this work, we extract DESI BGS-like samples from a realistic LSST mock galaxy catalogue, investigate the intrinsic galaxy size variations, simulate LSST-like images, and measure galaxy sizes to quantify the measurement uncertainties. Finally, we forecast the expected SNR for the dipole in joint DESI+LSST observations.

This paper is structured as follows. In Section~\ref{sec:theory}, we describe the theory of Doppler magnification, focusing on the dipole component of the two-point function, along with the corresponding estimator and its variance. In Section~\ref{sec:desi_lsst_galaxy}, we present the DESI-like sample selection from the \textsc{cosmoDC2} simulated catalogue \citep{Korytov_2019}, quantify the intrinsic galaxy size variations, and use simulated LSST-like images to assess measurement uncertainties, yielding the total per-bin uncertainty on the convergence used in our forecasts. In Section~\ref{sec:sigal_to_noise}, we present the forecast dipole amplitude and SNR in four redshift bins, and in Section~\ref{sec:conclusion} we discuss and conclude.

In what follows, we assume a Planck cosmology with $\Omega_c=0.261$, $\Omega_b=0.0490$, $h=0.677$, $\sigma_8=0.811$, $n_s=0.965$, and $\rm N_{eff}=3.046$ \citep{2020A&A...641A...6P}.

\section{Theory}
\label{sec:theory}

In this section, we outline the theoretical framework for Doppler magnification. We first define the number count contrast and lensing convergence, then describe how the Doppler magnification dipole can be extracted using cross-correlations. We then present expressions for the variance and measurement uncertainties.

\subsection{The Doppler magnification dipole}

\label{subsec:doppler_magnification_dipole}

The galaxy number count contrast $\Delta$ is defined as the fractional excess of galaxy counts in a direction $\bm{n}$ at redshift $z$ relative to the mean:
\begin{equation}
    \Delta(\bm{n}, z) =\frac{N(\bm{n}, z)- \left \langle  N(z)\right \rangle}{\left \langle  N(z)\right \rangle},
	\label{eq:delta0}
\end{equation}
where $N(\bm{n}, z)$ is the number of galaxies in a given volume element at observed redshift $z$ in the direction $\bm{n}$ (defined from the observer to the source), and $\langle  N \rangle$ is the mean number of galaxies over a sufficiently large volume. Regions where $\Delta>0$ are considered overdense, while regions with $\Delta<0$ are underdense. At linear order, the number count contrast can be approximated as a sum of density and velocity contributions (see e.g. \citealt{Andrianomena_2019}):
\begin{equation}
    \Delta(\bm{n}, z)\simeq b\delta -\frac{1}{\mathcal{H}}\partial _{r}(\bm{V} \cdot \bm{n}),
	\label{eq:Delta}
\end{equation}
where $b$ is the linear galaxy bias, $\delta$ is the matter density contrast, $\mathcal{H}\equiv a H$ is the conformal Hubble parameter, with $a$ the scale factor, $H$ is the expansion rate, $r$ is the comoving distance to the galaxy, and $\bm{V} \cdot \bm{n}$ is the line-of-sight component of the galaxy peculiar velocity field. Full expressions at linear order, including terms neglected above, can be found in \citet{Bonvin_2011,Challinor_2011}.

The lensing convergence $\kappa$ quantifies the fractional change (magnification or demagnification) in the apparent surface area of a background galaxy. It receives contributions from several effects: the gravitational convergence $\kappa_g$ (the standard weak gravitational lensing term), the Doppler convergence $\kappa_v$, the Sachs-Wolfe term $\kappa_{\text{\tiny SW}}$ and the Integrated Sachs-Wolfe term $\kappa_{\text{\tiny ISW}}$. The last two terms ($\kappa_{\text{\tiny SW}}$ and $\kappa_{\text{\tiny ISW}}$) are generally very small compared to $\kappa_g$ and $\kappa_v$ and are typically neglected \citep{Bacon_2014}. We therefore consider  $\kappa \simeq \kappa_g + \kappa_v$ for our analysis.

The Doppler convergence $\kappa_v$ arises from line-of-sight peculiar velocities shifting the observed redshift, and thus the apparent magnification can be influenced by both the foreground and background structures. Its contribution can be written as \citep{Bonvin_2008,Bacon_2014,Andrianomena_2019}
\begin{equation}
    \kappa_{v} = \left ( \frac{c}{r\mathcal{H}}-1 \right )\frac{\bm{V}\cdot \bm{n}}{c}.
	\label{eq:kappa_v}
\end{equation}
If a galaxy is moving towards the observer (i.e., $\bm{V}\cdot \bm{n} < 0$), then $\kappa_v < 0$, the galaxy is demagnified, and its observed angular size appears smaller. {This condition is satisfied at the low redshifts relevant for our analysis ($z\lesssim0.7$), where $r\mathcal{H}/c<1$.} Its flux, evaluated at fixed observed redshift $z_{\rm obs}$, is dimmer than if it had no peculiar motion. Conversely, for a galaxy moving away, it appears magnified. This differential effect produces a characteristic dipole pattern -- a spatial anisotropy pattern in the observed galaxy distribution around overdensities. 

Importantly, at low redshifts the Doppler magnification dominates over the gravitational lensing magnification. At $z \lesssim 0.45$, the Doppler term $\kappa_{v}$ exceeds the gravitational term $\kappa_{g}$, so that $\kappa_{g}$ can be safely neglected when extracting the dipole signal \citep{Bacon_2014, Bonvin_2017}. {In addition, $\kappa_g$ does not generate a purely (or predominantly) dipolar pattern along the line of sight. Galaxies behind an overdensity are lensed, whereas those in front are not. This also leads to monopole and even-multipole contributions. Since we extract the signal using a dipole estimator, as we discuss later in this Section, the resulting $\kappa_g$ contribution is relatively suppressed compared to the Doppler term.}

We can estimate $\kappa$ using a Bayesian approach \citep{Alsing_2015}. This method constructs a posterior distribution for $\kappa$ by incorporating prior knowledge of the intrinsic distribution of galaxy sizes and magnitudes, along with a model describing how lensing modifies these observables. The lensing convergence $\kappa$ affects the logarithm of the observed galaxy size -- here the size is defined as $\sqrt{\text{Area}}$, and $\lambda=\ln\sqrt{\text{Area}}$ -- and affects the magnitude $m$ linearly: \begin{equation}
    \lambda_{\text{obs}} = \lambda_{\text{int}} + \kappa, \\
    m_{\text{obs}} = m_{\text{int}} - q \kappa,
\end{equation}
where $ q = - 5 \log_{10} e \approx -2.17$, the subscript ``obs'' refers to the observed value and ``int'' refers to the galaxy’s intrinsic (unlensed) value.

{In this work, we will conservatively assume that $\kappa$ is inferred from galaxy sizes only. These are a natural by-product of galaxy shape measurement pipelines based on profile fitting, for example. See \citet{2023MNRAS.523.3649E} for a recent comparison of magnification measurements via size and flux observables, in this case from the DES survey.}

We adopt the half-light radius $R$ as a proxy for the galaxy size, such that $\lambda \approx \ln R$. The main factors that determine how well we can estimate $\kappa$ for each galaxy are the measurement noise in galaxy size and flux, and the intrinsic dispersion in galaxy sizes. {In practice, $\kappa$ can be inferred statistically from the observed sizes, magnitudes, and redshifts of galaxies by comparing their joint distribution to the observed distribution, rather than at the level of individual galaxies \citep{Alsing_2015, 2023MNRAS.523.3649E}.}

\cite{Bonvin_2017} and \cite{Andrianomena_2019} developed a formalism that optimally extracts the dipolar modulation induced by Doppler magnification from the cross-correlation between the galaxy number counts and convergence. The two-point cross-correlation function of $\Delta$ and $\kappa$ is given by 
\begin{equation}
    \xi ^{\Delta \kappa}=\left \langle \Delta (z,\bm{n})\kappa (z',\bm{n}') \right \rangle,
	\label{eq:xi}
\end{equation}
where $(z,\bm{n})$ and $(z',\bm{n}')$ denotes the redshift and line‑of‑sight direction at which $\Delta$ and $\kappa$ are evaluated, respectively.

One can use the following estimator to extract the dipole directly from survey data \citep{Bonvin_2017}:
\begin{equation}
    \hat{\xi}^{\Delta \kappa_{\nu}}_{\textup{dipole}}(d)=a_{\rm N}\sum_{ij}\Delta_{i}\,\kappa _{j}\cos\beta _{ij}\, \delta^K_{d_{ij},d}.
	\label{eq:dipole}
\end{equation}
Here, $a_{\rm N} = 3\,l_p^5/(4\pi\,V\,d^2)$ is a normalisation factor that depends on the cell size $l_p$ of the cubic grid which the fields are discretised onto for the correlation function measurement, and on the comoving volume $V$ of a redshift bin centred at $z$ with width $\Delta z$. The indices $i$ and $j$ denote voxels in the survey separated by comoving distance $d_{ij}$, which are summed over all $ij$ pairs. The quantity $\beta_{ij}$ is the angle between the separation vector $d_{ij}$ that connects $\Delta$ to $\kappa$ and the line-of-sight direction at the position where $\Delta$ is measured (see Figure~1 of \citealt{Andrianomena_2019}). The Kronecker delta function implements the binning in separation, $d$.

The dipole component of the two-point function for $\Delta$ and $\kappa_\nu$ may be computed theoretically as \citep{Andrianomena_2019}
\begin{equation}
    \xi^{\Delta \kappa_{\nu}}_{\textup{dipole}}\simeq \frac{1}{2\pi ^{2}}\left(1-\frac{1}{\mathcal{H}r}\right)
\left ( \lambda _1 + \frac{4}{15}\nu_1 \right ) P_1(\cos\beta),
\label{eq:xi_dipole}
\end{equation}
where $P_1$ is the Legendre polynomial of degree 1. {The approximate equality denotes use of the flat-sky approximation, and that we have neglected evolution between $z$ and $z^\prime$.}
The functions $\lambda_{1}$ and $\nu_{1}$ are given by
\begin{equation}
\begin{split}
\lambda_{1}(d,r,\beta) &= \int {\rm d}k\,k^2j_{1}(kd)\mathcal{G}(z',k)T^2(k)\mathcal{P}(k) \\
&\quad \times \left[ \frac{2b}{3\Omega_m} \left(\frac{k}{\mathcal{H}_0}\right)^2 D(z,k)\right. \left. + \right. \left.\frac{1}{3}\frac{k}{\mathcal{H}(z)} \mathcal{G}(z,k) \right],
\end{split}
\label{eq:lambda}
\end{equation}and
\begin{equation}
\nu_{1}(d,r,\beta) = \int {\rm d}k\,k^2j_1(kd)T^2(k)\mathcal{P}(k)\frac{k}{\mathcal{H}(z)} \mathcal{G}(z,k)\mathcal{G}(z',k),
\label{eq:nu}
\end{equation}
where $j_1$ is the spherical Bessel function with $\ell$ = 1; $\mathcal{P}(k)$ is the (dimensionful) primordial power spectrum, defined in terms of the primordial gravitational potential perturbation $\Psi_{\rm p}$ as $\left \langle \Psi_\mathrm{p}(k)\Psi_\mathrm{p} (k')\right \rangle = (2\pi)^3\, \mathcal{P}(k)\,\delta^D(k+k')$; $D(z,k)$ is the growth factor of the matter density perturbations, normalised as $D(a=1,k) = 1$; $T(k)$ is the $\Lambda$CDM matter transfer function, and
\begin{equation}
    \mathcal{G}(a,k) = \frac{2\,k\,\mathcal{H}}{3\,\mathcal{H}_0^2\,\Omega_{\rm m}}\,f\,D 
    \label{eq:G}
\end{equation}is the velocity growth factor, where $\Omega_{\rm m}$ is the fractional matter density at $z=0$, and $f\equiv{\rm d}\ln D/{\rm d}\ln a$ is the logarithmic growth rate.

Following Equation 7.8 from \cite{Dodelson_2020}, we write, in Equations~\ref{eq:lambda} and \ref{eq:nu},
\begin{equation}
    \mathcal{P}(k)T^2(k) = \frac{25}{9}\frac{\Omega_m^2H_0^4}{k^4c^4D^2(a)}\,P_{\rm m}(k,a),
    \label{eq:Pk_relation}
\end{equation}
where $P_{\rm m}(k,a)$ is the matter power spectrum. {Note that Equations~\ref{eq:G} and \ref{eq:Pk_relation} have made use of the continuity and Poisson equations, and so implicitly assume General Relativity as the underlying gravitational theory. The expressions would need to be modified in alternative theories of gravity. Since our forecasts use a fiducial cosmological model that assumes GR, it is appropriate to use these expressions. Further details regarding the effects of modifications to GR on the Doppler magnification dipole signal can be found in \citet{Andrianomena_2019}.}

\subsection{Variance of the dipole estimator}
\label{subsec:variance}

The covariance matrix for the Doppler magnification dipole estimator was also computed by \cite{Andrianomena_2019}, and is given by
\begin{align}
\textrm{cov}[\hat{\xi}^{\Delta \kappa_{\nu}}_{\textup{dipole}}] &= \Xi_1(z, d, d^\prime) + \Xi_2(z, d, d^\prime) + \Xi_3(z, d, d^\prime), \label{eq:cov}
\end{align}
where the individual terms are given by
\begin{align}
\Xi_1 &=\frac{9}{\text{V}}\left (1-\frac{1}{\mathcal{H}r} \right )^2 
\left (\frac{b^2}{5}+\frac{2bf}{7}+\frac{f^2}{9}\right )
\label{eq:cov1}\\
&\quad\quad \times f^2\frac{\mathcal{H}}{\pi^2} \int {\rm d}k\,P^2_{\rm m} (k,z)j_1(kd)j_1(kd') \notag \\
\Xi_2 & = \frac{9}{2}\sigma ^2_\kappa \frac{l^3_p}{V}\left (\frac{b^2}{3}+\frac{2bf}{5}+\frac{f^2}{7}\right ) 
\label{eq:cov2}\\
& \quad\quad \times \frac{1}{\pi ^2}\int {\rm d}k\,k^2 P^2_{\rm m}(k,z) j_1(kd)j_1(kd') \notag \\
\Xi_3 & = \frac{3}{4\pi }\frac{\sigma ^2_\kappa}{\bar{n}V} \Bigg(\frac{l_p}{d}\Bigg)^{2}\delta^K_{d,d'}.
\label{eq:cov3}
\end{align}
The first term is the contribution from the cosmic variance of $\Delta$, $\kappa$, and their cross-correlation. The second term quantifies the total variance associated with the size measurements $\sigma_\kappa$, which consists of the intrinsic size error and the measurement error. The third term, which depends on the mean number density of galaxies $\bar{n}$, arises from the impact of shot noise and also depends on $\sigma_\kappa$.

In this work, we use a simplified   $\sigma_\kappa$ estimation from galaxy sizes. Following \cite{Alsing_2015}, the transformation  of log-size under lensing (ignoring the magnitude effect) is given by $\langle \lambda \rangle \rightarrow \langle \lambda \rangle + \eta_\lambda \kappa$, where $\eta_\lambda = \partial \langle \lambda \rangle/\partial \kappa$ accounts for the boosting of galaxy sizes due to lensing and the shift of smaller sources across size selection boundaries. Assuming no selection effects, $\eta_\lambda=1$. The estimator for $\kappa$ is then
\begin{equation}
\hat{\kappa} = (\lambda - \langle \lambda \rangle) / \eta_\lambda.
\label{eq:kappa}
\end{equation}
The uncertainty on $\kappa$ is equal to the scatter in log-size, i.e., $\sigma_{\kappa}=\sigma_{\lambda}$. 
Using standard error propagation, the uncertainty in log-size is related to the uncertainty in radius via $\sigma_{\lambda}=\sigma_{R}/R$, thus $\sigma_{\kappa}=\sigma_{R}/R$.

\subsection{Uncertainty on the dipole measurement}

Accurate measurement of the Doppler magnification dipole requires both high-precision redshifts and robust size estimates. Spectroscopic surveys such as DESI minimise redshift uncertainties (e.g. compared with photometric surveys). Potential systematic biases on the reconstructed convergence field are not as well understood however, and could conceivably bias the dipole measurement or impact its SNR.

We assume that the observed convergence $\kappa_{\rm obs}$ is the sum of the `true' Doppler magnification convergence $\kappa_\nu$ and an additional noise term,
\begin{equation}
    \kappa_{\rm obs}=\kappa_{\nu} \, + \, \epsilon,
	\label{eq:kappa_obs}
\end{equation}
where $\epsilon$ is a random noise term which follows a Gaussian distribution with mean $b_{\kappa}$ (the measurement bias) and standard deviation $\sigma_{{\epsilon}}$,
\begin{equation}
\epsilon\sim\mathcal{N}(b_{\kappa},\,\sigma _{{\epsilon}}).
\end{equation}
{The measurement error, $\sigma _{\kappa}$, can be defined as
\begin{equation}
    \sigma _{\kappa}^2 = \langle (\kappa_{\rm obs} - \kappa_\nu)^2 \rangle.
\end{equation}}
$\sigma _{\kappa}$ includes two contributions: an intrinsic term $\sigma _{\kappa,\,\mathrm{intrinsic}}$ arising from variations in galaxy sizes and a measured term $\sigma _{\kappa,\,\mathrm{measured}}$ accounting for uncertainties in size measurement, which are assumed independent and so can be added in quadrature as
\begin{equation}
    \sigma _{\kappa} ^{2}= \sigma _{\kappa,\,\mathrm{intrinsic}} ^{2} + \sigma _{\kappa,\, \mathrm{measured}} ^{2}.
	\label{eq:sigma_kappa_total}
\end{equation}
The noise term $\epsilon$ therefore includes both intrinsic scatter and measurement uncertainty. The intrinsic component has zero mean by construction, but we permit the measured contribution to include a potential bias, $b_\kappa$. The measured contribution is calculated by taking the difference between the measured galaxy sizes and the true sizes, which we discuss later in Section~\ref{subsec:size_measurement}. The difference naturally includes the contribution from the bias. We assume that the measurement bias has no cosmological origin, and therefore does not correlate with the galaxy overdensity and does not contribute to the mean of the dipole estimator.

We define the effective $\sigma_{\kappa}$ as the weighted variance averaged over all selected galaxies in the survey,
\begin{equation}
    \sigma _{\kappa} ^{2}=\frac{\int {\rm d}m_r\,{\rm d}R\,{\rm d}z\;n(m_r,R,z)\:\sigma_\kappa ^{2}(m_r,R,z)}{\int {\rm d}m_r\,{\rm d}R\,{\rm d}z\;n(m_r,R,z)},
	\label{eq:sigma_kappa_eff}
\end{equation}
where $n(m_r,R,z)$ is defined as
\begin{equation}
    n(m_r,R,z)=\frac{{\rm d}N}{{\rm d}V\,{\rm d}m_r\,{\rm d}R},
	\label{eq:n}
\end{equation} ${\rm d}V$ is the comoving volume element, and $m_r$ is the $r$-band apparent magnitude. $\sigma_\kappa ^{2}(m_r,R,z)$ is the expected variance of the convergence measurement for a galaxy with $m_r$, $R$, $z$, which can be decomposed into the intrinsic scatter and the measurement error.

\section{Properties of the galaxy sample}
\label{sec:desi_lsst_galaxy}

In this section, we describe the selection of a galaxy sample to examine the number count distribution defined in Eq.~\ref{eq:n} and other properties. Galaxies are selected from the \textsc{cosmoDC2} catalogue based on the target selection criteria of the DESI Bright Galaxy Survey (BGS). \textsc{cosmoDC2} is a comprehensive simulated galaxy catalogue developed for LSST to support precision cosmology \citep{Korytov_2019}. It is based on the $(4.225\,\mathrm{Gpc})^3$ Outer Rim $N$-body simulation run and covers a sky area of 440 ${\rm deg}^2$ up to a redshift of $z = 3$. The catalogue reaches a depth of $28$ mag in the $r$ band, and includes a rich set of galaxy properties. Its construction combines semi-analytic galaxy formation modelling with empirical calibration. The semi-analytic component is based on the Galacticus model \citep{Benson_2012}, which models key baryonic processes such as gas cooling, star formation, and feedback. Additional empirical modelling is applied to reproduce observed relations, including the colour-magnitude and size-luminosity distributions \citep{Korytov_2019}.

Galaxy sizes in \textsc{cosmoDC2} are modelled by separately parameterizing disk and bulge half-light radii using the functional form of \cite{Zhang_2019}. These relations are calibrated to SDSS galaxies at $z \approx 0$ and evolve with redshift through a simple scaling that decreases the characteristic sizes by roughly a factor of two at $z = 1$. While the model ensures broad consistency with observed size-luminosity trends, the simulated galaxies are statistically realistic but physically simplified. The simulated galaxy size distribution does not necessarily reproduce that observed in real data in full; however, it provides a statistically consistent and physically plausible framework that is suited for this simulated analysis.

\subsection{Selection functions}
\label{subsec:selection_func}

Given that the dipole signal we aim to measure is intrinsically quite small, spectroscopic redshifts are preferred over photometric estimates due to their reduced uncertainties. We therefore propose using DESI spectroscopic data to construct the number count contrast field $\Delta$. To characterize the expected galaxy sample, we select galaxies from \textsc{cosmoDC2} according to the BGS target selection criteria.

While the DESI Peculiar Velocity (PV) Survey \citep{Douglass_2023,Saulder_2023} and the LOW-z Secondary Target Programme \citep{Ford_2023} were also considered, they are not suited to detecting the Doppler signal. The PV sample, although scientifically valuable for other peculiar velocity measurements (e.g. using the fundamental plane and the Tully-Fisher relations), has a much lower number density and covers only a limited redshift range ($z \approx 0.05 - 0.15$). Similarly, the LOW-z sample targets galaxies at $z < 0.03$, which falls entirely below our analysis threshold. Therefore we focus exclusively on the BGS sample.

To reproduce an equivalent BGS sample from the \textsc{cosmoDC2} catalogue, we adopt a magnitude cut of  $r < 19.5$, along with broad colour cuts of $-1 < (g-r) < 4$ and $-1 < (r-z) < 4$ \citep{Hahn_2023}. While official BGS criteria include fibre magnitude cuts to avoid imaging artifacts, these are not simulated in \textsc{cosmoDC2}. The BGS quality cut also requires at least one photometric observation in the $g, r, z$ bands, which the \textsc{cosmoDC2} catalogue inherently provides. We focus on the BGS Bright sample and exclude the fainter BGS Faint sample ($19.5 < r < 20.175$), as these galaxies might have low fibre flux and higher redshift uncertainty.

\begin{figure}
	\includegraphics[width=\columnwidth]{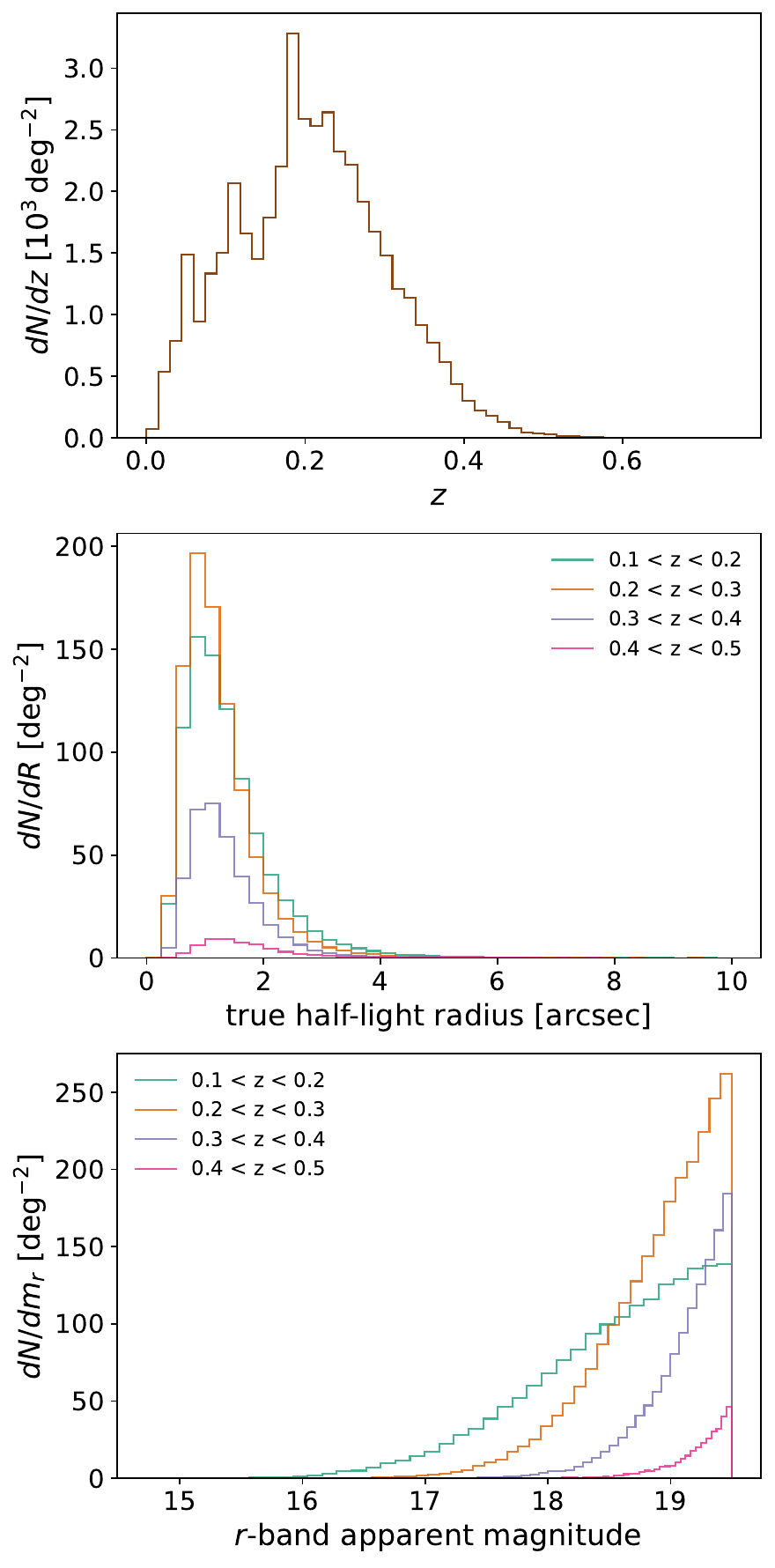}\caption{Distributions of galaxy properties from the BGS-selected \textsc{cosmoDC2} catalogue. Top: redshift distribution of the sample. Middle: true half-light radii (of the major axis) distribution in four redshift bins. Values above 10 arcsec are excluded from the plot for clarity. Bottom: apparent $r$-band magnitude distribution for the same redshift bins.}
    \label{fig:dN_distributions}
\end{figure}

Figure~\ref{fig:dN_distributions} presents the distributions of redshift, true half-light radius (of the major axis) and apparent $r$-band magnitude $m_r$ for the BGS-selected \textsc{cosmoDC2} galaxy sample. The redshift distribution $dN/dz$ peaks at $z \sim 0.2$, with a tail extending to $z\sim0.6$. The true half-light radius distribution ${\rm d}N/{\rm d}R$ shows that the majority of galaxies have angular sizes below 2 arcsec, with a steep decline at larger radii. We restrict the plotted range to galaxies with $0 < R < 10$ arcsec to exclude the very small number of extreme outliers with very large radii. The apparent $r$-band magnitude distribution ${\rm d}N/{\rm d}m_r$ shows a sharp increase in number counts toward the faint end, and peaks near the selection threshold at $r=19.5$. Overall, most sources in the BGS sample are concentrated at $z \sim 0.2$, with compact sizes of $R\sim 1$ arcsec, and relatively faint $m_r \sim 18\text{--}20$. For the forecast of the measurement of the Doppler magnification dipole, we choose four redshift bins $0.10\le z<0.20$; $0.20 \le z<0.30$; $0.30 \le z<0.40$; and $0.40 \le z<0.50$.

\subsection{The intrinsic $\sigma_\kappa$}
\label{sec:kappa_intrinsic}

\begin{figure} 
\includegraphics[width=\columnwidth]{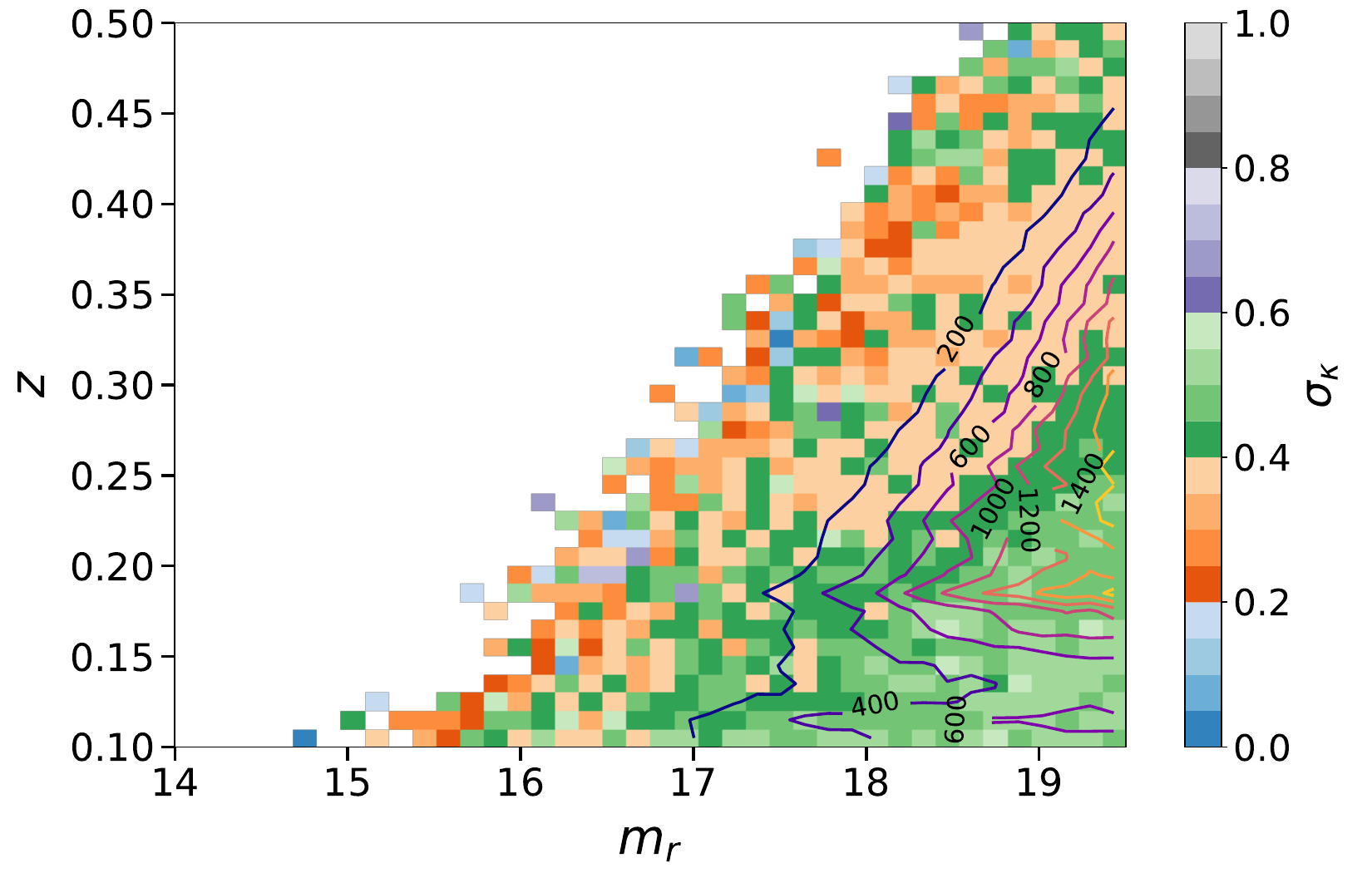}
    \caption{Distribution of $\sigma _{\kappa,\,\mathrm{intrinsic}}$ in bins of $m_r$ and $z$ for the DESI BGS-like simulated sample selected from the LSST \textsc{cosmoDC2} simulation. The colour scale represents $\sigma _{\kappa,\,\mathrm{intrinsic}}$ within each $(m_r,z)$ bin, ranging from 0 to 1, while the contours represent the number of galaxies selected from the catalogue, plotted in intervals of 200.
    }
    \label{fig:sigmaKappa_mr_z}
\end{figure}

\begin{figure} 
\includegraphics[width=\columnwidth]{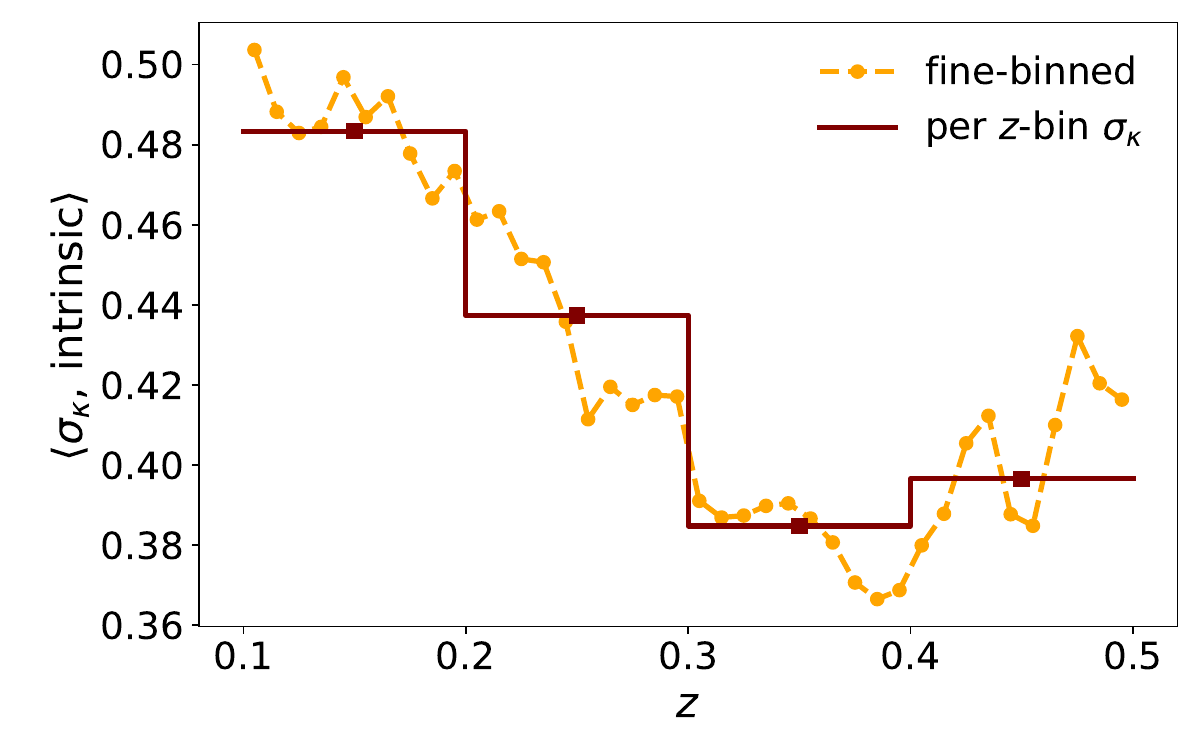}
    \caption{Count-weighted average $\langle \sigma_{\kappa,\,\mathrm{intrinsic}}\rangle$ in finely binned $z$ intervals and the chosen redshift bins. The orange dashed curve shows the values of the computed $\sigma_{\kappa,\,\mathrm{intrinsic}}$ from Figure~\ref{fig:sigmaKappa_mr_z} averaged over $m_r$. The maroon step function shows the average of $\sigma_{\kappa,\,\mathrm{intrinsic}}$ within the chosen  redshift bins, with the square markers at the midpoints of each redshift bin.}
    \label{fig:sigmaKappa_z_finned}
\end{figure}

The convergence field can be estimated from the galaxy sizes and magnitude measurements \citep{Alsing_2015}, and the associated error on the convergence, $\sigma_\kappa$, is a key factor influencing the overall noise level in the dipole signal. The total uncertainty in $\kappa$ arises from two sources: an intrinsic component due to the inherent variation in galaxy sizes, and measurement-related errors (see Eq.~\ref{eq:kappa_obs}). By analysing the size distribution of galaxies selected from the simulation, we aim to characterize the expected intrinsic size variation, and thereby anticipate the noise level on the convergence field.

The intrinsic contribution $\sigma_{\kappa,\,\mathrm{intrinsic}}$, as defined in Equation~\ref{eq:sigma_kappa_eff}, is the population-weighted average of $\sigma_\kappa^{2}(m_r,R,z)$ across the survey. Integrating over $m_r$, $R$, and $z$, weighted by the number density from Equation~\ref{eq:n}, is equivalent to an average over the full selected galaxy population. Therefore, we need to find the expected intrinsic scatter for each ($m_r$, $R$, $z$) bin in our selected population. 

From the catalogue, we select a total of 276,578 BGS-like galaxies. To evaluate $\sigma_{\kappa,\,\mathrm{intrinsic}}^{2}(m_r,z)$, we bin the galaxy population in the space of apparent $r$-band magnitude and redshift. Within each grid cell, we compute the $\sigma_\kappa^{2}$ as $(\text{std}(R)/\bar{R})^2$, where $\bar{R}$ is the true half-light radius. Figure~\ref{fig:sigmaKappa_mr_z} shows the resulting values of $\sigma_\kappa^{2}$ across all grid cells, with the contour indicating the number density of the galaxies. For each redshift bin in $0.10\leq z\leq 0.50$, we compute a count-weighted average of $\sigma_\kappa^{2}$ over all $(m_r,z)$ cells within that bin, yielding an effective value of $\sigma_{\kappa,\,\mathrm{intrinsic}}$ of the galaxy population at each redshift bin. The results are summarized in Table~\ref{tab:sigma_kappa} and visualized in Figure~\ref{fig:sigmaKappa_z_finned}. Figure~\ref{fig:sigmaKappa_z_finned} shows the finely binned $\langle \sigma_{\kappa,\,\mathrm{intrinsic}}\rangle$ as a function of $z$, averaged over $m_r$, along with the same quantity computed in the four broad redshift bins as reported in Table~\ref{tab:sigma_kappa}.

With the 3D-HST+CANDELS sample, \cite{van_2014} found an intrinsic scatter $\sigma_{\log{\rm R_{{\rm eff}}}} \simeq 0.1\text{--}0.14$ for early-type galaxies and $\simeq 0.16\text{--}0.19$ for late-type galaxies over $0<z<3$, where we have taken these values to be in dex ($\log_{10}$). At higher redshifts using JWST NIRCam, \cite{Allen_2025} measured an intrinsic scatter of $\sigma_{\log{\rm R_{{\rm eff}}}}=0.188 \pm 0.007$ over $3 < z < 9$, within $1\sigma$ agreement with values at lower redshift from \cite{van_2014}, where $R_{{\rm eff}}$ denotes the effective radius. These results suggest that the intrinsic scatter has little to no evolution with redshift. The values found from the simulated catalogue are broadly consistent with these results.\footnote{For reference, we approximate $\sigma_{\log{\rm R_{{\rm eff}}}} = 0.1 (0.2)$ as $\sigma_\kappa \approx 0.23 (0.48)$.}

\begingroup
\begin{table}
\setlength{\tabcolsep}{5pt} 
\centering
\begin{tabular}{cccccc}
\hline
$z$ & $N_{\mathrm{obj}}$ & $\sigma_{\kappa,\,\mathrm{measured}}$ & $\sigma_{\kappa,\,\mathrm{intrinsic}}$ & $\sigma_{\kappa,\,\mathrm{total}}$ & {$b_{\kappa}$} \\
\hline
$[0.10, 0.20]$ & 2311 & 0.027 & 0.48 & 0.48 &  {0.0056} \\
$[0.20, 0.30]$ & 4038 & 0.032 & 0.44 & 0.44 &  {0.0037} \\
$[0.30, 0.40]$ & 2443 & 0.032 & 0.39 & 0.39 & 0.0055 \\
$[0.40, 0.50]$ & 666  & 0.029 & 0.40 & 0.40 & {0.0089} \\
\hline
\end{tabular}
\caption{The estimates of the intrinsic and measured $\sigma_{\kappa}$ values in four redshift intervals used in our forecasts. For each redshift bin we list the number of simulated galaxies $N_{\mathrm{obj}}$ from the set of 10,000 randomly selected BGS-like galaxies (see Section~\ref{sec:image_simulation}), the measurement contribution $\sigma_{\kappa,\,\mathrm{measured}}$ from the image-simulation pipeline (as described in Section~\ref{subsec:size_measurement}), the intrinsic size dispersion $\sigma_{\kappa,\,\mathrm{intrinsic}}$ (Equation~\ref{eq:sigma_kappa_eff}), their quadrature sum $\sigma_{\kappa,\,\mathrm{total}}$ which is used in our analysis, and the mean bias $b_{\kappa}$ for reference.}
\label{tab:sigma_kappa}
\end{table}
\endgroup

\subsection{Image simulations}
\label{sec:image_simulation}

We employ the modular image simulation package \textsc{GalSim}\footnote{\url{https://github.com/GalSim- developers/GalSim}} \citep{rowe2015galsim} to generate galaxy images using simple profiles and properties (such as S\'ersic index $n$, half-light radius $R$) as well as observational effects such as the point spread function (PSF), detector characteristics and sky background noise. Subsequently, we measure the sizes of the galaxies in these simulated images to assess biases in size recovery across a variety of DESI BGS-like galaxies. This image simulation is set up to mimic the expected data quality of the LSST Year 1 (Y1) observations.

For simulated galaxy images, we randomly selected 10,000 DESI BGS-like galaxies from the \textsc{cosmoDC2} catalogue. For each selected galaxy, we generate images using true half-light radius, apparent $r$-band magnitude and redshift from the catalogue. All other input parameters are held fixed except for the random realisation of image CCD noise. The galaxy flux for each image is set by its apparent magnitude, instrumental zero point, and exposure time $t_{\rm exp}$ as
\begin{equation}
F_{\rm gal} = \frac{t_{\rm exp}}{\rm gain} \cdot 10^{-0.4 \cdot (m - Z)},
\label{eq:galaxy_flux}
\end{equation}
where $F_{\rm gal}$ is the flux of the galaxy, $t_{\rm exp}$ is the exposure time, $m$ is the apparent magnitude of the galaxy, and $Z$ is the photometric zero point.
We adopt $Z = 28.36$ mag, corresponding to the LSST instrumental zero point for a 1-second exposure at unit gain, and a detector gain of 1.6 electrons per ADU.\footnote{\url{https://smtn-002.lsst.io}} The single-visit exposure time is set to 30 seconds \citep{lssthandbook_2009}.


To reduce complexity, we model each galaxy as a single-component system with a S\'ersic surface brightness profile and a fixed S\'ersic index of $n=1.5$. This is a simplifying assumption regarding the S\'ersic index: early-type elliptical galaxies or bulge components typically have $n\sim4$ and late-type spiral galaxies or disk components have $n\sim1$ \citep[e.g.,][]{Shen_2003,Lange_2015}. Given that most galaxies in our target redshift range ($z \sim 0.1 - 0.5$) are late-type, we consider $n=1.5$ a reasonable choice. As we discuss later in Section~\ref{subsec:size_measurement}, we find the measured size error to be much smaller than the intrinsic size error. We also test the extreme case of $n=4$, and find the same conclusion.

The image pixel scale is set to 0.2 arcsec/pixel, matching the LSST CCD design \citep{Ivezi_2019}. Each side of each image stamp has a pixel number equal to 25 times the galaxy’s half-light radius divided by the pixel scale. The simulated galaxies are isolated, without nearby stars, galaxies, or otherwise. We do not account for ellipticity in the galaxy models, as the summary statistics we use are averaged over a large number of galaxies. This simplifying assumption effectively treats all the galaxies as viewed face-on. The PSF model includes both atmospheric and optical components, which are convolved to form a total PSF. The atmospheric PSF is modeled with a Kolmogorov profile and a fiducial seeing of 0.7 arcsec \citep{Ivezi_2019}. The optical PSF accounts for telescope effects: a wavelength-to-diameter ratio of 0.017, a linear obscuration of 40$\%$ of the primary mirror diameter, and four support struts with thickness of 0.03. The struts are oriented at an angle of 10 degrees. Optical aberrations are also included. Both the optical PSF and the aberration parameters follow the \texttt{GalSim} LSST example configuration.\footnote{\url{https://github.com/GalSim-developers/GalSim/blob/releases/2.7/examples/lsst.yaml}} 

The simulated noise characteristics include both Poisson (photon) noise and Gaussian read noise. A uniform sky background of 18,000 ADU per arcsec$^2$ is added to each image, and the Gaussian read noise is set to 3.4 rms per pixel per read, following the values in the \texttt{GalSim} LSST example configuration. The noise is applied using independent random seeds, ensuring unique noise realisations. To approximate the noise level for the LSST-Y1 coadded image, the noise is scaled by a factor of $1/\sqrt{82.5}$. This is based on the LSST baseline of $\sim825$ visits per field over ten years \citep{PSTN-051}, and reflects that coadding $N$ independent exposures reduces the noise level by a factor of $\sqrt{N}$. The full 10-year LSST data quality will benefit from deeper coadds, so the corresponding $\sigma_{\kappa,\,\mathrm{measured}}$ is expected to be slightly smaller than in this Y1 simulation. A few examples of the simulated galaxy images are presented in Appendix~\ref{app:galaxy_images} (Figure~\ref{fig:app:sim_galaxies}).

For each simulated galaxy, we generated and archived three types of images: the final galaxy image that is convolved with PSF and noise, the corresponding PSF-only image (i.e. containing only the PSF model, without galaxy or noise), and the noise rms image. These images are the inputs required for size measurements using \textsc{galight} (see Section \ref{subsec:size_measurement}). In real observing scenarios, these components would be provided as estimates based on the full images.

\begin{figure}
    \centering
        \includegraphics[width=0.98\linewidth]{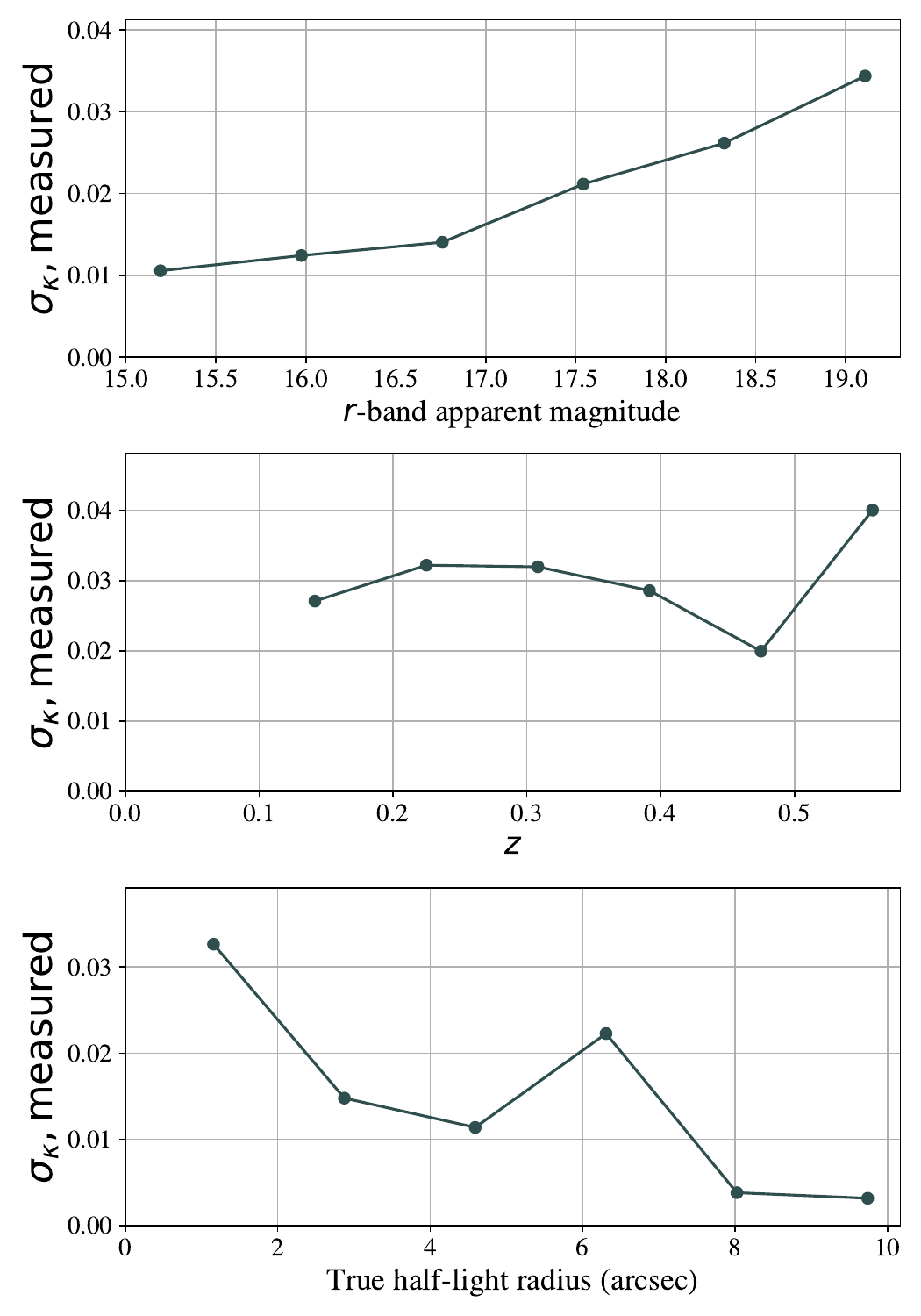}
    \caption{Measurement uncertainty in galaxy half-light radius, $\sigma_{\kappa,\, \mathrm{measured}}$, with varied $r$-band apparent magnitude (top), redshift (middle), and lensed half-light radius (bottom), for a sample of 10,000 galaxies from the \textsc{cosmoDC2} catalogue selected to match DESI BGS-like criteria. Galaxies with $z\leq0.05$ are excluded from the analysis.}
    \label{fig:sigmaR_vs_all}
    
\end{figure}

\subsection{Size measurement}
\label{subsec:size_measurement}

We used \textsc{galight} \citep{Ding_2020} to estimate the apparent size of each galaxy from the simulated images. \textsc{galight} is an open-source \texttt{python} package for two-dimensional image modelling, built on the image modelling capabilities of \textsc{lenstronomy} \citep{Birrer_2021}. It provides image processing, PSF handling, and 2D profile fitting for galaxy structural components. In our implementation, the fitting is performed individually for each image. The fitting process incorporates both the PSF and the noise rms map, as saved during the image generation step. We use the shallow fitting level in the fitting process and choose the particle swarm optimization (PSO) algorithm \citep{Kennedy_PSO} to maximise the model likelihood.

To characterise the accuracy of our galaxy size measurements, we quantified the measurement error as the mean squared difference between the best-fit and true half-light radii, averaged over the selected 10,000 galaxies:
\begin{equation}
\sigma_{\kappa,\, \mathrm{measured}}^2 = \Big \langle \big(R_{\rm measured} - R_{\rm true}\big)^2/ R_{\rm true}^2\Big \rangle.
\label{eq:residual_R}
\end{equation}
We present the measurement uncertainty of galaxy sizes (half-light radius), $\sigma_{\kappa,\, \mathrm{measured}}$, across bins of $r$-band apparent magnitude, redshift, and true galaxy half-light radius in Figure~\ref{fig:sigmaR_vs_all} to check the dependence of measurement uncertainty on key physical properties. The results show that $\sigma_{\kappa,\, \mathrm{measured}}$ slightly increases towards fainter magnitudes, varies modestly with redshift, and tends to decrease with larger true galaxy sizes. Note that our measurements assume a simplified single-component model for all galaxies, consistent with the image simulation inputs. The spike features in the lower two panels of Figure~\ref{fig:sigmaR_vs_all} are interpreted as random fluctuations in bins with small numbers of simulated galaxies.

We then compute the uncertainty per redshift bin. The results are shown in Table~\ref{tab:sigma_kappa}. Across all redshift bins, the measurement scatter in galaxy sizes is much smaller than the intrinsic size dispersion. Measurement uncertainties are therefore subdominant, contributing only a small fraction of the total $\sigma_{\kappa}$. {As discussed earlier, the measurement bias $b_\kappa$ has been accounted for in the calculation of $\sigma_{\kappa,\, \mathrm{measured}}$.
We computed the mean $b_\kappa$ as $\langle R_{\rm measured} - R_{\rm true} \rangle$. The results are shown in Table~\ref{tab:sigma_kappa}. Here we only computed $b_\kappa$ for reference.}

We have assumed a particular image simulation setup with specific assumptions for image systematic effects, and also a particular galaxy profile fitting approach. All of these aspects could differ in the real LSST images, and the actual analysis will use different profile measurement tools for example. Nevertheless, this is intended to represent a broadly realistic measurement process, with many of the expected complications, and $\sigma_{\kappa,\,\mathrm{measured}}$ clearly remains sub-dominant to the intrinsic size scatter at all times. 

In addition, the fact that the estimator correlates two different fields from different experiments and extracts only a dipole provides further protection against uncorrelated systematics, such as PSF modelling errors, fluctuations in the galaxy number density, or observational variations (e.g. changes in seeing, sky background, blending, or detector response), which are expected to mostly correlate out. In this sense, the Doppler magnification dipole should be quite a robust observable. This should be compared with other measures of peculiar velocities, such as Tully-Fisher/Fundamental Plane, which do require a careful treatment of systematic errors \citep[e.g.,][]{Johnson_2014}.

\section{Forecasts}
\label{sec:sigal_to_noise}

In this section, we quantify the amplitude and detectability of the Doppler magnification dipole for a DESI+LSST-like survey. We adopt the redshift-dependent bias and number density values given in Table 1 of \cite{Andrianomena_2019}, and construct the covariance assuming an effective overlap sky area of $\sim 5,900$ ${\rm deg}^2$ \citep{Olsen_2018}. Note that this is an approximate estimate of the overlap, which is likely smaller in practice. However, in the near future, the 4MOST survey will provide an additional overlapping spectroscopic galaxy redshift sample for LSST galaxies \citep{Jong_2019}. We assume the observed cubic grid has cell size $l_p=4\,{\rm Mpc}/h$.

\begin{figure}	\includegraphics[width=\columnwidth]{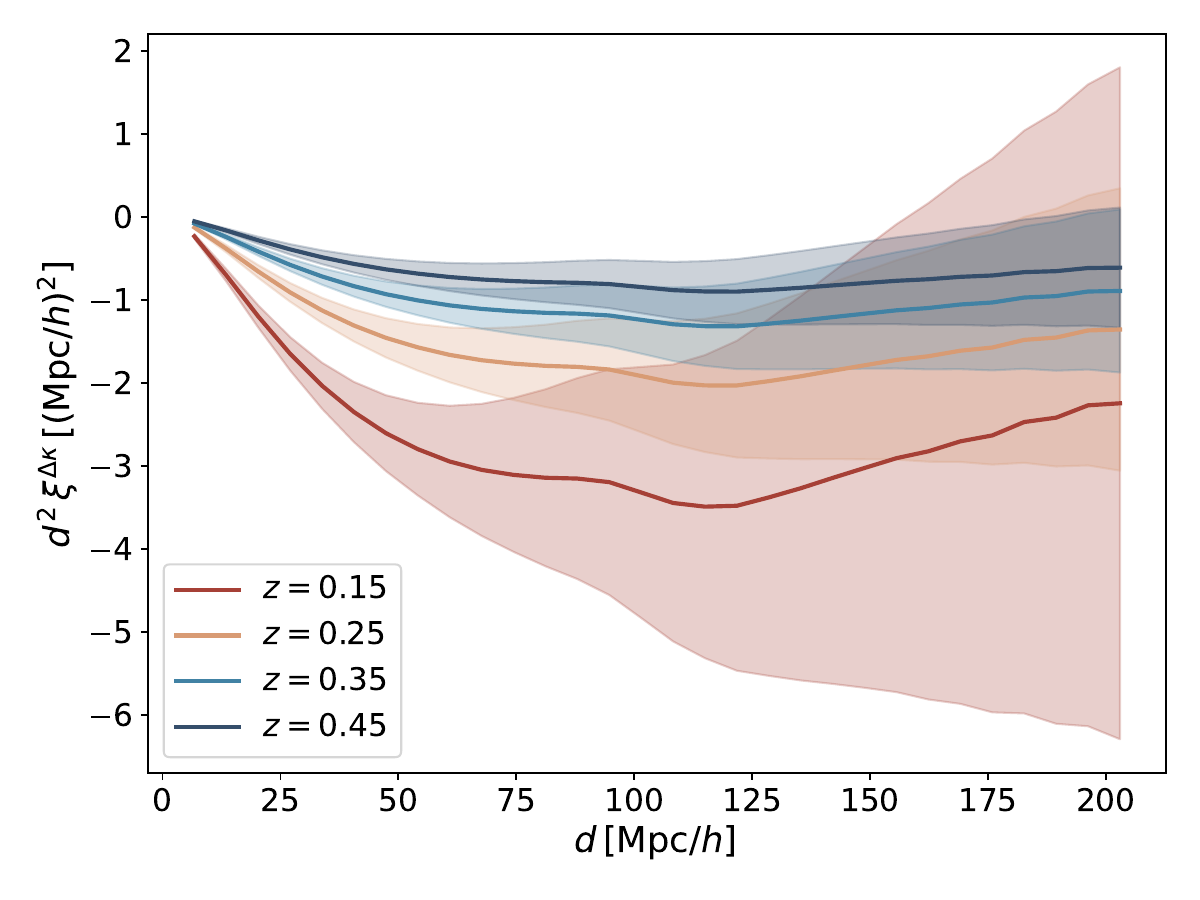}
    \caption{Predicted Doppler magnification dipole $\xi^{\Delta\kappa}$ for a DESI+LSST-like survey in four redshift bins. The solid curves show the amplitude at different redshifts, with shaded regions representing 1-$\sigma$ uncertainties from the calculated covariance matrix (Equation~\ref{eq:cov}), computed with the size-noise values $\sigma_{\kappa,\mathrm{total}}$ for each bin from Table~\ref{tab:sigma_kappa}.}
    \label{fig:xi}
\end{figure}

Figure~\ref{fig:xi} presents the predicted Doppler magnification dipole $\xi^{\Delta\kappa}(d)$ for the selected DESI+LSST-like sample in four redshift bins, $z
\sim [0.15,\,0.25,\,0.35,\,0.45]$, as defined in Section~\ref{subsec:selection_func}. The amplitude of the dipole is calculated according to Equation~\ref{eq:xi_dipole}. In the figure, the error band is the standard deviation, which is the square root of the diagonal of the covariance matrix (Equation~\ref{eq:cov}). The separation $d$ ranges from 10 Mpc to 310 Mpc, with a uniform bin width of 10 Mpc. For each redshift bin we adopt the corresponding per-bin total galaxy size variation noise values $\sigma_{\kappa,\,\mathrm{total}}$ from Table~\ref{tab:sigma_kappa}. The value of $\sigma_{\kappa,\,\mathrm{total}}$ is computed following Equation~\ref{eq:sigma_kappa_total}, where the intrinsic and measurement components are combined in quadrature. 

The amplitude of $\xi^{\Delta\kappa}(d)$ is negative across the full range. It drops rapidly for $d \simeq 10\text{--}50\,{\rm Mpc}/h$, reaches a broad minimum at $d \simeq 100\text{--}125\,{\rm Mpc}/h$, and then turns over, becoming less negative toward larger separations. At fixed separation, the (absolute) amplitude decreases with redshift. The lowest-$z$ bin has the largest (most negative) amplitude. The 1-$\sigma$ error bars widen toward large separations for all bins and are broader at low $z$ than at high $z$. Several factors contribute, including the fact that fewer galaxy pairs exist at large separations, and the comoving volume is smaller at lower $z$.

Following \cite{Bonvin_2017}, we define the dipole SNR as
\begin{equation}
\frac{S}{N}(d) = \frac{\left|\langle \hat{\xi}^{\Delta \kappa_{\nu}}_{\textup{dipole}} \rangle(d)\right|}{\sqrt{\textrm{cov}[\hat{\xi}^{\Delta \kappa _{\nu }}_{\textup{dipole}}](d,d)}},
\label{eq:sn_def}
\end{equation}
where $\langle \hat{\xi}_{\rm dipole} \rangle(d)$ is the mean dipole signal at separation $d$, and $\textrm{cov}[\xi^{\Delta \kappa_{\nu }}](d,d)$ is the diagonal element of the covariance matrix at $d$, as described in Equation~\ref{eq:cov}.

\begin{figure}	\includegraphics[width=\columnwidth]{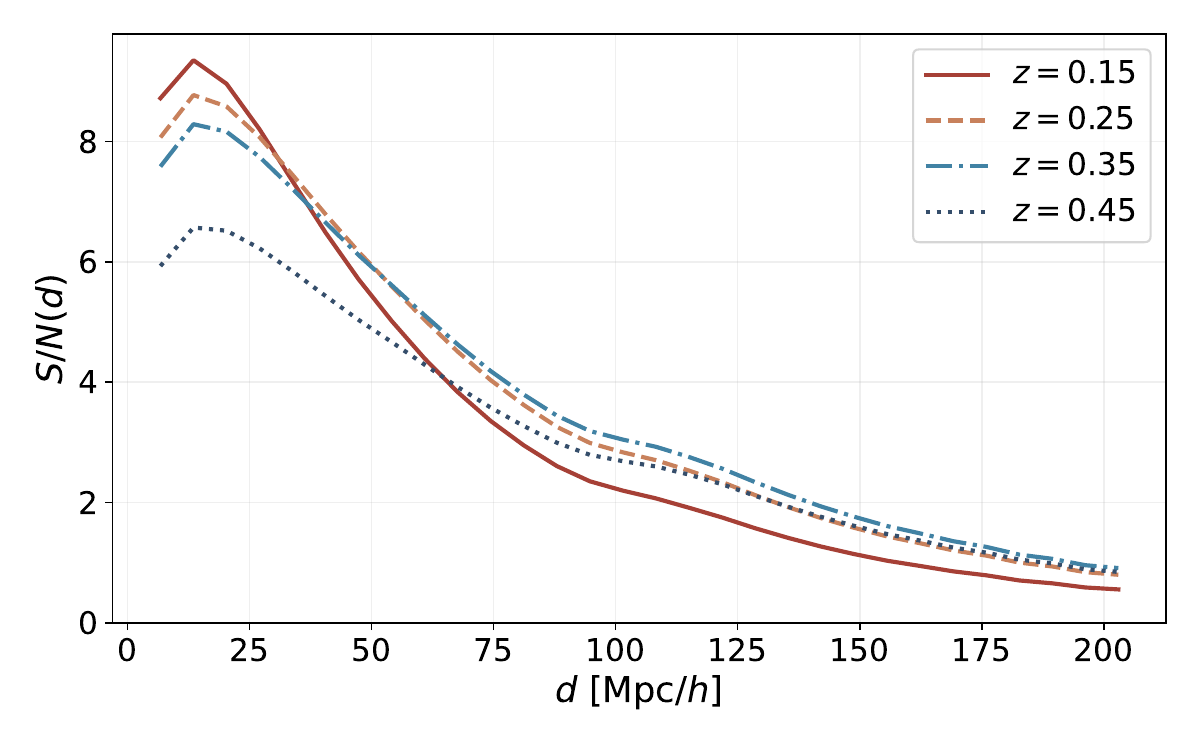}
    \caption{Signal-to-noise ratios of the Doppler magnification dipole for a DESI+LSST-like survey in four redshift bins. The S/N peaks at small separations ($\sim 10\text{--}25$ ${\rm Mpc}/h$) and slowly declines with increasing separation.}
    \label{fig:snr}
\end{figure}

Figure~\ref{fig:snr} shows the dipole SNR in four redshift bins. For all of the bins, the curves peak at small separations, reaching $S/N$ $\simeq 8$ at $z=0.15$ and $z=0.25$, and slightly lower values ($S/N\simeq7$) at $z=0.35$, near $d \sim 10\text{--}25$ ${\rm Mpc}/h$. The $S/N$ then declines smoothly with increasing $d$. At the smallest separations, the lowest $z$ bin has the highest SNR because of the larger dipole amplitude at low redshift. Beyond $d\sim30 \, {\rm Mpc}/h$, this trend reverses slightly, likely because larger comoving volumes reduce sample variance and $\sigma_{\kappa}$ declines modestly with redshift, even as the mean number density decreases.

The values remain at a moderate level ($S/N \approx4\text{--}5$) out to $d\approx 50\text{--}75$ ${\rm Mpc}/h$. A very shallow increase is visible near $d \approx 90\text{--}125 \, {\rm Mpc}/h$. By $d \approx 200 \, {\rm Mpc}/h$, the SNR values for all bins decrease to $\approx1$. We find that the total S/N per $z$ bin are 12.98, 12.97, 12.66 and 10.41 for $0.1 \le z < 0.2$, $0.2 \le z < 0.3$, $0.3 \le z < 0.4$ and $0.4 \le z < 0.5$, respectively. We also compute the SNR using the dipole expression in \cite{Bonvin_2017} instead of the expression in \cite{Andrianomena_2019}, and find very similar results. These values are lower than those reported by \cite{Andrianomena_2019}, mostly because we have a smaller sky area, corresponding to the overlap region between LSST and DESI.

\begin{figure*}
    \includegraphics[width=2.0\columnwidth]{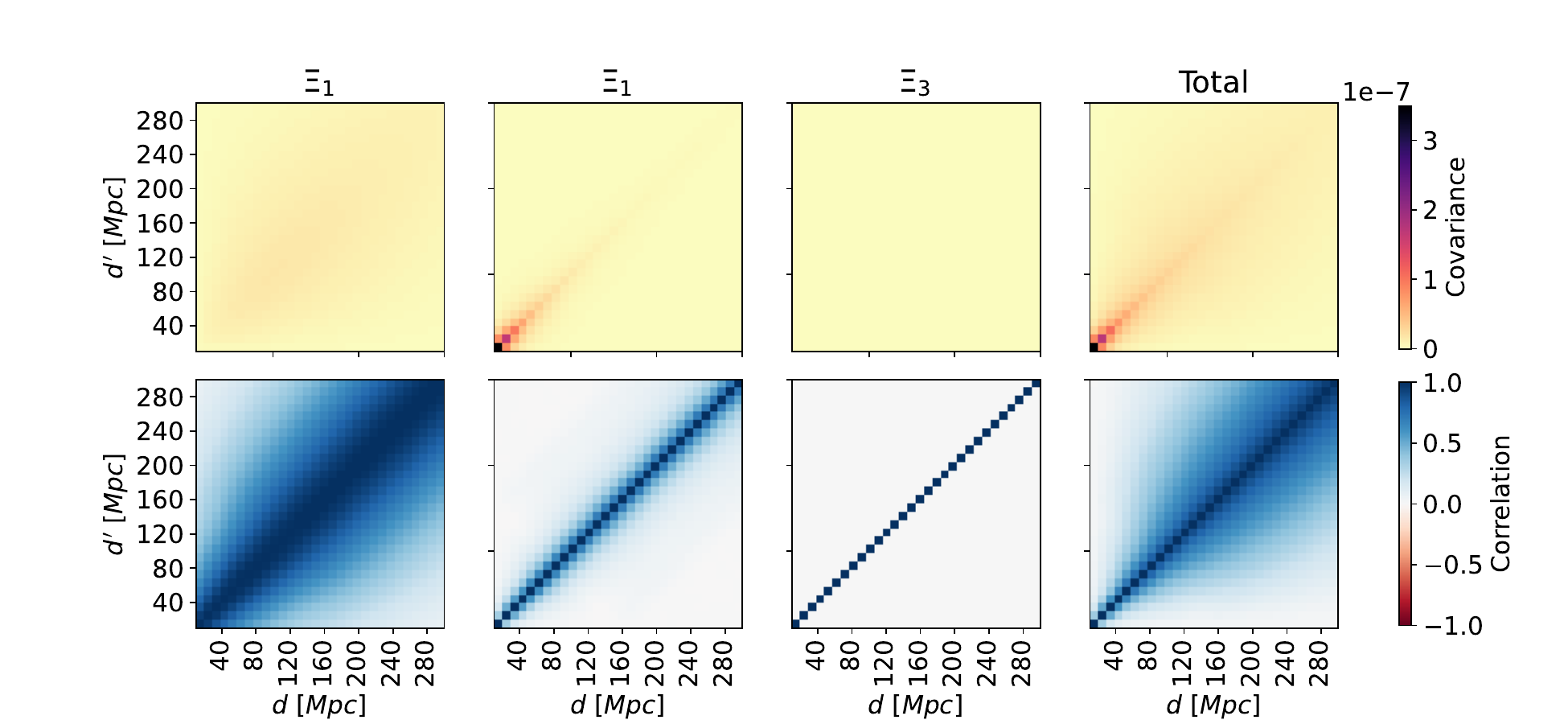}
    \caption{Top panels: Covariance matrices of the Doppler magnification dipole for $0.1\le z<0.2$. The $\Xi_1$, $\Xi_2$, $\Xi_3$, and ``Total'' panels correspond to the components specified in Equation~\ref{eq:cov}. Bottom panels: The corresponding correlation matrices at $0.1\le z<0.2$ (Equation~\ref{eq:rho_ij}). Axes denote separations $d$ and $d^\prime$, with bins of width 10 Mpc and axis ticks shown every 40 Mpc for clarity. Each row shares a single colour bar.}
    \label{fig:covar_corr}
\end{figure*}

For reference, we show the covariance matrix and the correlation matrix for $\xi^{\Delta\kappa_\nu}$ at $0.1\le z<0.2$ in Figure~\ref{fig:covar_corr}. Both the covariance and correlation matrices are plotted separately for each of the three terms and for their sum (Equation~\ref{eq:cov}). The correlation matrix is defined as
\begin{equation}
    \rho_{ij} = \frac{C_{ij}}{\sqrt{C_{ii} \cdot C_{jj}}},
    \label{eq:rho_ij}
\end{equation}
where $C_{ij}$ represents the covariance between measurements at separations $d_i$ and $d_j$.

The covariance is largest at the small separations and decreases towards larger $d$. Once normalized, it is more obvious that the bins at large separations are strongly correlated. $\Xi_1$ has strong correlations across a wide range of scales and corresponds to the cosmic variance contribution to the signal. $\Xi_2$ is correlated only for neighbouring separation bins. $\Xi_3$ is purely diagonal, consistent with being shot noise that does not correlate between different separations. At $d\lesssim40\,{\rm Mpc}$, $\Xi_2$ provides the largest contribution to the variance, reflecting the impact of $\sigma_{\kappa}$. Beyond this scale, $\Xi_1$ (cosmic variance) dominates, while $\Xi_3$ remains small at all separations. 

Across all redshift bins, $\Xi_3$ contributes the least among all three components of the covariance (See Figure~\ref{fig:app:cov}). Towards higher redshifts, the cosmic variance term $\Xi_1$ contributes less as expected. At the highest two redshift bins, $\Xi_2$ completely dominates over $\Xi_1$ at all scales. Additionally, any reduction in $\sigma_{\kappa}$ mainly helps at small separations. Increasing the survey volume through a larger sky area is useful for improving the SNR in the $0.1 \le z < 0.2$ bin, but has little impact in the higher redshift bins.

\section{Conclusions}
\label{sec:conclusion}

We have presented a simulated measurement process and forecasts for detecting the Doppler magnification dipole with a joint DESI+LSST analysis, combining DESI BGS-like galaxy number densities and biases with the convergence field that would be recovered assuming an LSST-informed size-noise model. Using \textsc{cosmoDC2} to define DESI BGS-like samples, we estimated the per-galaxy size-noise $\sigma_{\kappa}$ per redshift bin. We separated the total $\sigma_{\kappa}$ into intrinsic size dispersion and measurement error. The intrinsic size dispersion was estimated directly from the true half-light radii of the simulated galaxies, averaged over the DESI BGS-like population in each redshift bin. We used \textsc{GalSim} and \textsc{Galight} to generate LSST Y1-quality images and subsequently test the size recovery error. Across all bins, the measurement scatter is well below the intrinsic dispersion under our simplified assumptions, which include single-component S\'ersic profiles (fixed $n$), no ellipticity, and fixed LSST-like PSF/noise configurations. We forecasted detectability by propagating the estimated $\sigma_{\kappa}$ into the covariance for a DESI-area survey, using the Doppler magnification dipole estimator from \cite{Andrianomena_2019}.

The predicted dipole is negative across all separations, steepens around $d \simeq 10 \text{--} 50\,{\rm Mpc}/h$, reaches a broad minimum near $d\simeq100\text{--}125\,{\rm Mpc}/h$, and becomes less negative toward larger $d$. The amplitude decreases with redshift over $z \simeq 0.15\text{--}0.45$. The total S/N per $z$ bin is 12.98, 12.97, 12.66 and 10.41 for $0.1\leq z<0.2$, $0.2\leq z< 0.3$, $0.3 \leq z <0.4$ and $0.4\leq z < 0.5$, respectively. The corresponding per-bin significance peaks at $S/N \simeq 8\text{--}9$ around $d \simeq 10\text{--}25\,{\rm Mpc}/h$, and falls to $S/N \simeq 1\text{--}2$ by $d \simeq 200\,{\rm Mpc}/h$. Taken together, these results indicate that the dipole should be measured with high significance with this combination of surveys.

From the covariance decomposition, we find that in the lowest redshift bin, the term proportional to $\sigma_{\kappa}$ dominates the variance at small separations and correlates only neighbouring bins. At larger separations, the cosmic variance term became dominant. The purely diagonal shot-noise term contributes the least at all scales. In higher redshift bins, the cosmic variance term decreases significantly, leaving the $\sigma_{\kappa}$-proportional term as the dominant contribution across nearly all separations. The shot noise-related component remains negligible in all redshift bins. Overall, the total variance is governed by the size noise and cosmic variance, with their relative importance depending on scale and redshift. An increase in the survey volume via a larger survey area would help to enhance the S/N in the lowest redshift bin, but not for the others.

Several caveats should be noted. The simulations presented here are based on a specific LSST-like image setup, and employ a particular galaxy profile fitting approach. Variations in observing conditions, actual sky coverage, and the use of different measurement pipelines could affect the recovered size distributions. In addition, the galaxy sizes in \textsc{cosmoDC2}, which are based on a semi-analytic model and empirical calibration, may not fully reproduce the size distributions that will be observed in real data. 

Another caveat is that our galaxy size uncertainty estimates were derived from \textsc{GalSim} simulations of isolated galaxies and thus do not account for the impact of blending. Blending, defined as having at least $1\%$ flux contribution from overlapping neighbours, is expected to affect $\sim 60\%$ of galaxies in 10-year full-depth LSST images \citep{Sanchez_2021}, and about $20\text{--}30\%$ of objects may be unrecognised blends \citep{Troxel_2023}. This reduces the usable galaxy sample and measurement precision. Ignoring such overlapping sources could therefore underestimate the size measurement uncertainties.

As one final note, we also neglected the possibility that other imaging surveys could be used, e.g. the Dark Energy Survey \citep{2016MNRAS.460.1270D} or DESI's own legacy imaging surveys \citep{2019AJ....157..168D}. The latter in particular is, by definition, well-matched with the DESI sample.

In this work, we demonstrated that current and near-future wide-field surveys should be capable of detecting the Doppler magnification dipole. The detectability depends on the effective number density and overlapping volume of the spectroscopic and imaging samples, with performance further conditioned on whether practical requirements (e.g., wide sky coverage, well-modelled PSF and controlled noise) are met. In the DESI+LSST set-up we modelled, intrinsic galaxy-size scatter dominates the size measurement noise, resulting in an adequate S/N for detection at low redshift and a detectable signal across all redshift bins considered.

\section*{Acknowledgements}

We are grateful to A.~Nicola for useful comments and discussions.
This result is part of a project that has received funding from the European Research Council (ERC) under the European Union's Horizon 2020 research and innovation programme (Grant agreement No. 948764; P.~Bull). I.~Ye acknowledges support from the China Scholarship Council (grant no. 202208060318). P.~Bull acknowledges support from STFC Grants ST/T000341/1 and ST/X002624/1. CC is supported by STFC grant ST/X000931/1.

\section*{Data Availability}

The code used in this study will be available upon request.

\balance

\bibliographystyle{mnras}
\bibliography{dopplermag} 

\appendix
\section{Example simulated galaxy images}
\label{app:galaxy_images}

\begin{figure*}
\centering
\includegraphics[width=1.95\columnwidth]{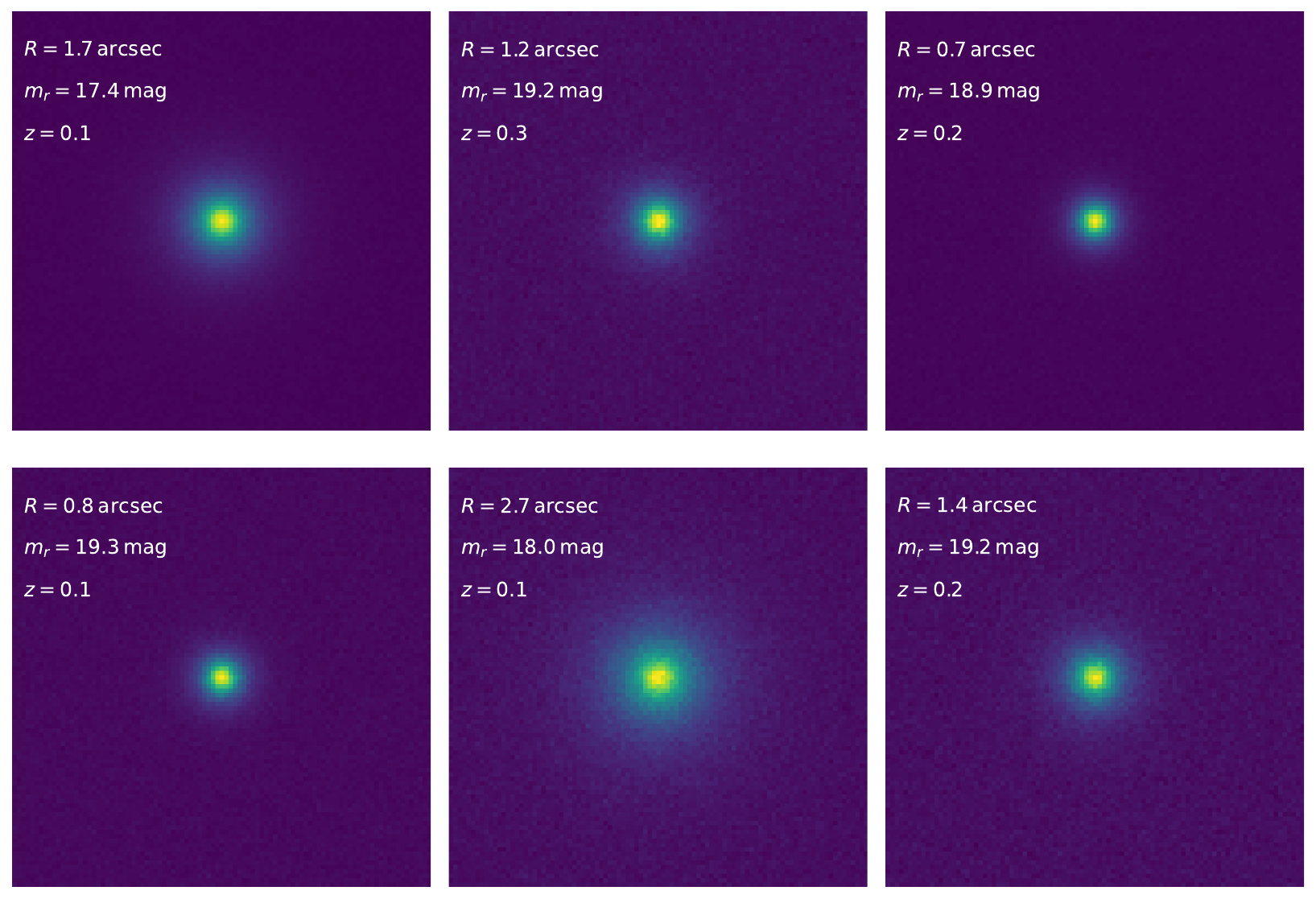}
    \caption{Example simulated LSST galaxy images generated using \textsc{GalSim}. The six panels show galaxies with varying intrinsic half-light radii ($R$), $r$-band apparent magnitude ($m_r$), and $z$ drawn from 
    the BGS sample. The corresponding physical parameters are indicated within each panel. Each galaxy was modelled as a S\'ersic profile, and convolved with a atmospheric and optical PSF, including aberrations and detector noise 
    appropriate for LSST observing conditions. The images have a pixel size of $(0.2\, \rm{arcsec})^2$, with 95 pixels on each side.
    }
    \label{fig:app:sim_galaxies}
\end{figure*}

To illustrate the simulated sample used in this analysis, Figure~\ref{fig:app:sim_galaxies} displays a random selection of six example galaxy images generated with \textsc{GalSim}. These images were drawn from the mock sample based on DESI BGS, and incorporate realistic photometric, atmospheric, 
and instrumental effects consistent with anticipated LSST observing conditions. Each galaxy was modelled as a S\'ersic profile with index $n = 1.5$, convolved with a composite PSF  that includes atmospheric Kolmogorov turbulence and optical aberrations (defocus, astigmatism, and coma). Detector noise was added using the expected LSST gain, sky level, and read noise parameters.

\section{Covariance decomposition}
\label{app:covariance}

\begin{figure*}
\centering
\includegraphics[width=1.90\columnwidth]{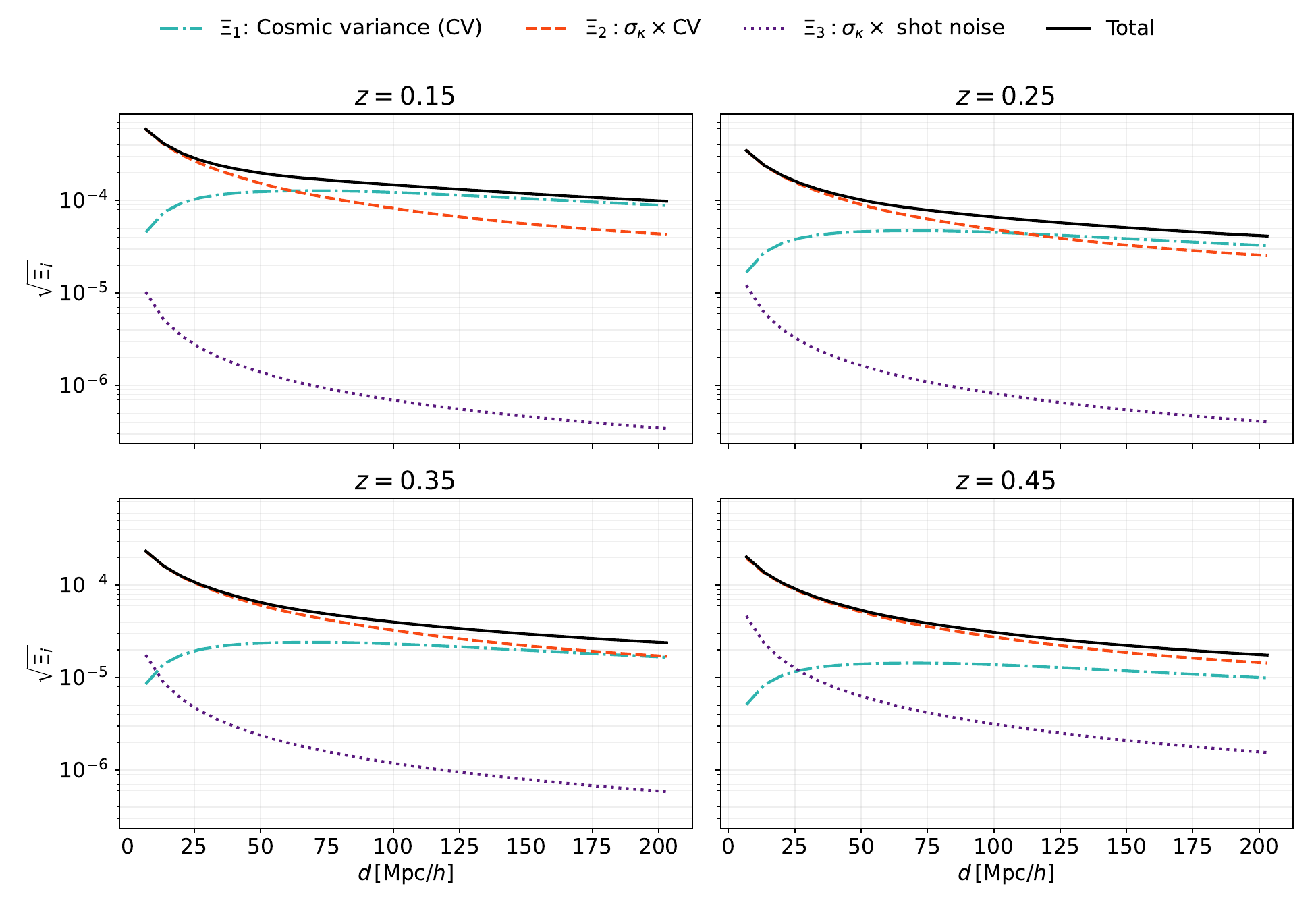}
    \caption{Square root of the diagonal of the covariance contributions $\sqrt{\Xi_i(d)}$ and their total sum, where $i=1,2,3$ (see Equation~\ref{eq:cov}--\ref{eq:cov3}), plotted as a function of $d$ for the four redshift bins used in this work. 
    }
    \label{fig:app:cov}
\end{figure*}

In Section~\ref{sec:sigal_to_noise}, we summarised how different components of the covariance contribute to the total uncertainty of the dipole estimator. For further clarity, Figure~\ref{fig:app:cov} shows the redshift and scale dependence of the contributions $\Xi_i$. These correspond to the cosmic variance term, the $\sigma_\kappa$ dependent term, and the shot noise plus $\sigma_\kappa$ dependent term (see Equations~\ref{eq:cov1}--\ref{eq:cov3}). 

As shown in the figure, $\Xi_2$ dominates the cosmic variance at small separations in all bins. $\Xi_1$ becomes increasingly important at large separations in the lower redshift bins. At higher redshifts, the amplitude of $\Xi_1$ decreases substantially, leaving $\Xi_2$ to contribute the most across most scales. The shot noise term $\Xi_3$ is negligible throughout. 

\bsp	
\label{lastpage}
\end{document}